\documentclass{optica-article}

\journal{opticajournal} 

\articletype{Research Article}

\usepackage{lineno}
\usepackage{mathtools,braket,verbatim}

\definecolor{teal}{RGB}{26,157,150}

\newcommand{\br}{\mathbf{r}}
\newcommand{\cE}{\mathcal{E}}
\newcommand{\cH}{\mathcal{H}}
\newcommand{\cP}{\mathcal{P}}
\newcommand\hmu{\hat{\mu}}
\newcommand\hnu{\hat{\nu}}
\newcommand\hsigma{\hat{\sigma}}
\newcommand\hF{\hat{F}}
\newcommand\hC{\hat{C}}
\newcommand\hV{\hat{V}}
\newcommand\ha{\hat{a}}
\newcommand\hb{\hat{b}}
\newcommand\hc{\hat{c}}
\newcommand\tgamma{\tilde{\gamma}}
\newcommand{\imag}{\mathrm{i}}

\usepackage[capitalize]{cleveref}
\crefname{section}{Sec.}{Secs.}
\Crefname{section}{Section}{Sections}

\crefrangelabelformat{section}{#3#1#4--#5\crefstripprefix{#1}{#2}#6}
\crefrangelabelformat{figure}{#3#1#4--#5\crefstripprefix{#1}{#2}#6}
\crefrangelabelformat{equation}{(#3#1#4--#5\crefstripprefix{#1}{#2}#6)}

\crefmultiformat{equation}%
{\edef\crefstripprefixinfo{#1}Eqs.~(#2#1#3}%
{,#2\crefstripprefix{\crefstripprefixinfo}{#1}#3)}%
{,#2\crefstripprefix{\crefstripprefixinfo}{#1}#3}%
{,#2\crefstripprefix{\crefstripprefixinfo}{#1}#3)}

\begin{document}

\title{Beyond critical coupling: optimal design considerations for spontaneous four-wave mixing in microring resonators}

\author{Joseph M. Lukens,\authormark{1,2,*} Karthik V. Myilswamy,\authormark{1,$\dagger$} Alexander Miloshevsky,\authormark{3} and Hsuan-Hao Lu\authormark{2}}
\address{\authormark{1}Elmore Family School of Electrical and Computer Engineering and Purdue Quantum Science and Engineering Institute, Purdue University, West Lafayette, Indiana 47907, USA\\
\authormark{2}Quantum Information Science Section, Oak Ridge National Laboratory, Oak Ridge, Tennessee 37831, USA\\
\authormark{3}Electric Power Board of Chattanooga, Chattanooga, Tennessee 37402, USA
}

\vspace{1em}
\email{\authormark{*}jlukens@purdue.edu} 
\address{\authormark{$\dagger$}Currently with National Institute of Standards and Technology, Boulder, Colorado 80305, USA}


\begin{abstract*}
We present a self-contained analytical model for biphoton generation in microring resonators. Encompassing both all-pass and add-drop geometries, identical and distinct pump and biphoton coupling coefficients, and continuous-wave and pulsed pumping, our interaction-picture-based approach reveals time-frequency biphoton correlations while also predicting absolute generation rates. Under continuous-wave excitation, we find critical coupling of both the pump and biphoton to maximize the rate of single photons extracted from the microring, whereas critical coupling of the pump but overcoupling of the biphoton maximize the two-photon rate.
Under pulsed pumping, overcoupling of both pump and biphoton (to different degrees) maximizes photon extraction probabilities, albeit under a tradeoff with spectral factorability that we quantify via parameter scans over a range of coupler pairings. 
As a whole, our formalism should prove valuable for the practical design of integrated photon sources, merging a flexible and intuitive biphoton-centric depiction with quantitative predictions closely tied to experimental parameters.
\end{abstract*}

\section{Introduction}
\label{sec:intro}
Microring resonators form a staple of the integrated photonics toolkit~\cite{Bogaerts2012, Dai2014}, leveraged for filters~\cite{Little1997, Barwicz2004, Xia2007}, wavelength-division multiplexers~\cite{Dahlem2011, Orcutt2012}, pulse shapers~\cite{Agarwal2006, Khan2010, Wang2015b, Cohen2024a, Wu2025b}, modulators~\cite{Xu2005, Xu2007, Wang2018d}, and frequency comb generation~\cite{Xue2015,Shen2020, Wu2025}.
In the quantum domain, microring-based spontaneous four-wave mixing (SFWM) has arguably cemented itself as  \emph{the} standard for efficient photon-pair generation in CMOS photonic integrated circuits~\cite{Kues2019}, supporting a variety of upgrades ranging from interferometric couplers~\cite{Vernon2017, Liu2020, Burridge2023,  Borghi2024} to cascades of multiple tunable sources~\cite{Liscidini2019, Clementi2023, Borghi2023}. 
Analogous to bulk cavities, the microring's power 
stems from field enhancement---the ability to trap light inside the device for many round trips.  As quantified by the cavity (or intrinsic) quality factor $Q_c = \omega_0/\gamma_c$ 
with $\omega_0$ the resonance frequency and $\gamma_c$ the dissipation rate, intrinsic quality factors $Q_c>10^6$ have become common in many popular materials, including but not limited to AlGaAs ($\sim$3.5$\times10^6$~\cite{Xie2020}), LiNbO$_3$ ($\sim$2.9$\times10^7$~\cite{Zhu2024}), and SiN ($\sim$10$^8$~\cite{Puckett2021}).

Yet such high $Q_c$ values are of practical use for SFWM only when the microring is coupled to a larger photonic circuit. For pump light must enter the ring---and generated photons then be extracted therefrom---in order to exploit the microring's field enhancement in the outside world. Quantitatively, with coupling rates $\gamma_a$ and $\gamma_b$ to bus $a$ and drop $b$ waveguides, the total quality factor $Q$ is no longer $Q_c$ but reduced to $Q=\omega_0/(\gamma_a+\gamma_b+\gamma_c)$, 
reflecting an inherent tradeoff in microring design: intracavity field enhancement is maximized by \emph{eliminating} coupling, whereas photon extraction efficiency is maximized by \emph{increasing} coupling without bound. Historically, this tradeoff has typically been balanced by critical coupling, i.e., $\gamma_a=\gamma_b+\gamma_c$, for which all incoming on-resonant light is coupled into the microring~\cite{Bogaerts2012}. 
While maximizing field enhancement of the pump, critical coupling is not necessarily optimal for biphoton generation, since the total number of photon pairs extracted from the ring depends on additional outcoupling factors.

Such nuances have been recognized in a handful of previous works. For example, \cite{Vernon2016} notes the distinction between maximizing total photon rate compared to heralding efficiency, while \cite{Vernon2017} and experimental successors~\cite{Tison2017, Liu2020, Burridge2023,  Borghi2024} have focused on device geometries with distinct coupling rates for the pump and biphoton, thereby permitting separate coupling optimization for the input and output. Yet although the theoretical tools for coupling-coefficient optimization in microring-based SFWM exist in multiple sources~\cite{Helt2010, Vernon2015, Chembo2016, Liu2024, Fontaine2025}, no prior work has applied them to this specific problem: \emph{for a given geometry and pump power, what coupling coefficients maximize the rate of extracted photon pairs?}

In this article, we analytically derive and numerically analyze optimal coupling coefficients in microring-based SFWM, under a variety of configurations: (i)~all-pass versus add-drop geometries, (ii)~identical versus distinct pump and biphoton coupling coefficients, and (iii)~continuous-wave (CW) versus pulsed pumping. Combining the conceptual simplicity of interaction-picture-based formalisms~\cite{Scholz2009,Chen2011} with careful tracking of proportionality constants to predict absolute rates~\cite{Chembo2016, Liu2024}, our self-contained model outlines design principles for microring sources tailored to experimentalists. Interestingly, of the twelve concrete quantities optimized below---i.e., one- and two-photon rates under six combined geometry and pump configurations---critical coupling is optimal for the pump in only three of them, and for the biphoton in only two.
Given a fixed ring and average pump power, 
absolute two-photon rates are maximized by the add-drop geometry with distinct pump-biphoton coupling, 
the all-pass case with identical coupling is a close second, and the add-drop case with identical coupling finishes last. 
Overall, our findings address overlooked gaps in the existing quantum photonics design repertoire and are poised for immediate applications in microring-based quantum information processing circuits.

\section{Theoretical model}
\label{sec:theory}

\begin{figure}[tb!]
\includegraphics[width=\textwidth]{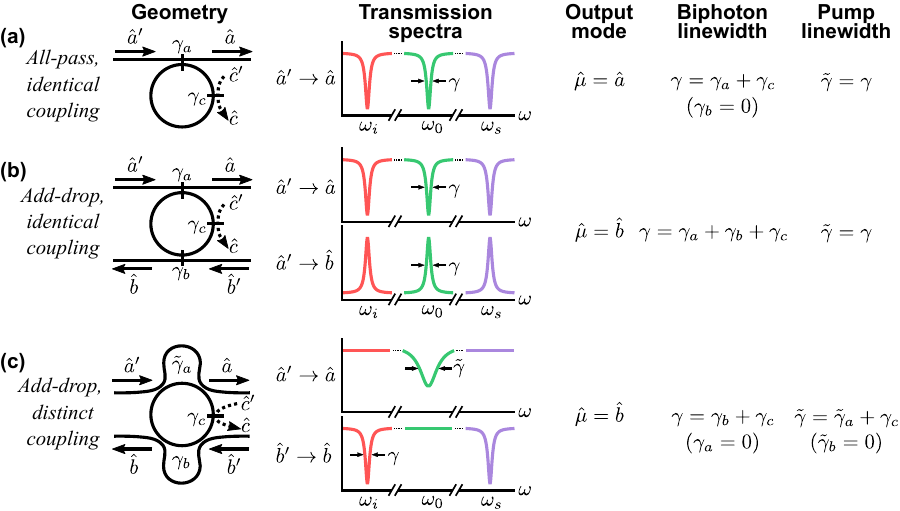}
\caption{Basic microring SFWM configurations considered. (a)~All-pass geometry with identical pump-biphoton coupling rates. (b)~Add-drop geometry with identical pump-biphoton coupling rates. (c)~Add-drop geometry with distinct pump-biphoton coupling rates. Each case can be uniquely identified by the output mode $\hmu\in\{\ha,\hb\}$ and triplet of coupling conditions defining the linewidth---namely, $\gamma=\gamma_a+\gamma_b+\gamma_c$ for the biphoton and $\tgamma=\tgamma_a+\tgamma_b+\tgamma_c$ for the pump.}
\label{fig:setup}
\end{figure}

The general microring configurations under study are depicted in Fig.~\ref{fig:setup}, namely: (a) an all-pass geometry with identical pump and biphoton coupling rates, (b) an add-drop geometry with identical pump and biphoton coupling, and (c) an add-drop geomotry with distinct pump and biphoton rates. The pump enters through mode $\ha'$ in all cases, while the signal and idler photons are collected in $\ha$ (all-pass) or $\hb$ (add-drop);  intracavity losses are modeled via a point coupler with virtual modes $\{\hc',\hc\}$.
The plotted resonances for idler $\omega_i$, pump $\omega_0$, and signal $\omega_s$ frequency bins show each geometry's ideal interaction with the outside world. All cases can be subsumed under a single formalism by letting $\hmu\in\{\ha,\hb\}$ denote the output mode and $(\tgamma_a,\tgamma_b,\tgamma_c)$ the coupling rates felt by the pump, as distinct from the rates $(\gamma_a,\gamma_b,\gamma_c)$ of the biphoton; the specific coupling conditions and relationships for each design are summarized in the rightmost three columns of Fig.~\ref{fig:setup}.

Mathematically, we begin with slowly varying time-domain annihilation operators $\tilde{\nu}(t)$ associated with each cavity coupling mechanism $\nu\in\{a,b,c\}$ (reserving ``hats''  for the Fourier transforms $\hnu$ introduced below).
The operators external to the cavity are normalized such that $\tilde{\nu}^\dagger(t)\tilde{\nu}(t)$ denotes photon flux in units of s$^{-1}$. Inside the cavity, the annihilation operator $\tilde{C}(t)$ is normalized such that $\tilde{C}^\dagger(t)\tilde{C}(t)$ denotes the total number of photons in the cavity at time $t$---i.e., $\tilde{C}^\dagger(t)\tilde{C}(t)$ is dimensionless. Under these definitions, standard coupled mode theory~\cite{Haus1984, Little1997, Liu2024} adapted to our conventions specifies the dynamical equations
\begin{equation}
\label{eq:coupling}
\tilde{\nu}(t) = -\tilde{\nu}'(t) + \sqrt{\gamma_\nu}\tilde{C}(t) \qquad ; \qquad \nu\in\{a,b,c\}
\end{equation}
for $\tilde{\nu}'(t)$ the mode impinging upon the coupler and $\tilde{\nu}(t)$ the mode exiting the coupler. Although we specifically adopt a phase convention with the output $\pi$-phase shifted from the input~\cite{Haus1984,Liu2024}, arbitrary phase references can be implemented without any change to our findings~\cite{Little1997}. The fictitious modes $\tilde{c}'(t)$ and $\tilde{c}(t)$ account for cavity loss in a lumped fashion---an approximation which holds in the envisioned scenario of low loss per round trip (high $Q_c$).

The equation of motion for the intracavity mode follows directly as
\begin{equation}
\label{eq:intracavity}
\frac{d}{dt}\tilde{C}(t) = -\frac{\gamma}{2}\tilde{C}(t) + \sum_\nu\sqrt{\gamma_\nu}\tilde{\nu}'(t),
\end{equation}
where $\gamma=\gamma_a+\gamma_b+\gamma_c$. Fourier-transforming the above via the definitions $\hnu(\Omega)=\frac{1}{\sqrt{2\pi}}\int dt\,\tilde{\nu}(t)e^{\imag\Omega t}$ and $\hC(\Omega) = \frac{1}{\sqrt{2\pi}}\int dt\,\tilde{C}(t)e^{\imag\Omega t}$ returns the solution
\begin{equation}
\label{eq:Csoln}
\hC(\Omega) = \frac{1}{\frac{\gamma}{2}-\imag\Omega}\sum_\nu \sqrt{\gamma_\nu}\hnu'(\Omega) = \frac{1}{\frac{\gamma}{2}+\imag\Omega}\sum_\nu \sqrt{\gamma_\nu}\hnu(\Omega),
\end{equation}
expressed in terms of either the input (primed) or output (unprimed) operators, all of which satisfy the commutation relations $[\hmu(\Omega),\hsigma^\dagger(\Omega')]=\delta(\Omega-\Omega')\delta_{\mu\sigma}$ for $\mu,\sigma\in\{a,a',b,b',c,c'\}$. (Throughout this paper any summation over a Greek index includes all three coupling pathways; i.e., $\sum_\nu=\sum_{\nu\in\{a,b,c\}}$ is implied.) Both expressions in \cref{eq:Csoln} possess the standard complex Lorentzian form $\left(\frac{\gamma}{2}\mp \imag\Omega\right)^{-1}$, with the choice of sign depending on whether input or output operators are considered. Incidentally, this sign has physical significance in ensuring causality of the pulsed pumped wavepacket derived in  \cref{eq:wavepacketFinal} below.

Focusing first on the output operator representation and noting that the above equations hold separately for both signal $s$ and idler $i$, we arrive at the positive-frequency \emph{intracavity} field operators $\hF_s^{(+)}(z,t)$ and $\hF_i^{(+)}(z,t)$:
\begin{equation}
\begin{split}
\label{eq:intraFlux}
\hF_s^{(+)}(z,t) & \equiv \sqrt{\frac{v_g}{L}}e^{\imag(k_s z - \omega_s t)} \tilde{C}_s(t) = \sqrt{\frac{v_g}{2\pi L}} e^{\imag(k_s z - \omega_s t)} \sum_\nu\sqrt{\gamma_\nu}\int_{-\infty}^\infty \frac{d\Omega}{\frac{\gamma}{2}+\imag\Omega}e^{-\imag\Omega t}\hnu_s(\Omega), \\
\hF_i^{(+)}(z,t) & \equiv \sqrt{\frac{v_g}{L}}e^{\imag(k_i z - \omega_i t)} \tilde{C}_i(t) =  \sqrt{\frac{v_g}{2\pi L}} e^{\imag(k_i z - \omega_i t)} \sum_\nu\sqrt{\gamma_\nu}\int_{-\infty}^\infty \frac{d\Omega}{\frac{\gamma}{2}+\imag\Omega} e^{-\imag\Omega t}\hnu_i(\Omega),
\end{split}
\end{equation}
where we have assumed negligible group-velocity dispersion (GVD) and taken $\omega_s$ and $\omega_i$ on resonance such that $k_sL$ and $k_iL$ equal a multiple of $2\pi$~\cite{noteScholz2009Chen2011}.
Under these conditions, $\hF_s^{(-)}(z,t)\hF_s^{(+)}(z,t)$ and $\hF_i^{(-)}(z,t)\hF_i^{(+)}(z,t)$ return the photon flux associated with each resonance at position $z\in(0,L)$ inside the cavity of the length $L$. 
The associated electric-field operators can be obtained by multiplying the above expressions by the slowly varying factor $\sqrt{\frac{\hbar\omega_0}{2\epsilon_0cn_0 S}}$, with $n_0$ the effective index at $\omega_0$ and $S$ the effective area of the transverse mode~\cite{Blow1990}. Intuitively, the normalization in \cref{eq:intraFlux} follows from the fact that the dimensionless photon fields $\tilde{C}_{s,i}(t)$ travel the entire ring in a time $L/v_g$ (length divided by group velocity); 
hence, the rate of photons passing any given intracavity point must equal $\frac{v_g}{L}\tilde{C}^\dagger_{s,i}(t)\tilde{C}_{s,i}(t)$.


The \emph{extracavity} signal-idler fields at either the through ($\mu=a$) or drop ($\mu=b$) port can be written generically as 
\begin{equation}
\begin{split}
\label{eq:extraFlux}
\hV_s^{(+)}(t) & \equiv \tilde{\mu}_s(t) = \frac{1}{\sqrt{2\pi}} \int_{-\infty}^\infty d\Omega\,e^{-\imag\Omega t} \hmu_s(\Omega), \\
\hV_i^{(+)}(t) & \equiv \tilde{\mu}_i(t) = \frac{1}{\sqrt{2\pi}} \int_{-\infty}^\infty d\Omega\,e^{-\imag\Omega t} \hmu_i(\Omega),
\end{split}
\end{equation}
again normalized such that $\hV_s^{(-)}(t)\hV_s^{(+)}(t)$ and $\hV_i^{(-)}(t)\hV_i^{(+)}(t)$ define photon flux.

The fields for the pump centered at frequency $\omega_0$ and propagation constant $k_0$ proceed along similar lines, with three main modifications: (i)~the coupling rates assume ``tilde'' values $\tgamma_\nu$ to allow them to be distinct from those of the signal-idler; (ii)~the intracavity field is written in terms of input (primed) operators [the first equality in \cref{eq:Csoln}]; and (iii)~the input field $\ha_p'(\Omega)$ is taken classically as a complex number with all other input fields neglected. Under CW pumping, we can take $\ha_p'(\Omega)=\sqrt{\frac{2\pi \cP}{\hbar\omega_0}}\delta(\Omega)$ for average power $\cP$; under pulsed pumping, we define $\ha_p'(\Omega) = \sqrt{\frac{\cE}{\hbar\omega_0}} A_p(\Omega)$ with $\cE$ the pulse energy and $\int d\Omega |A_p(\Omega)|^2=1$. Therefore the extracavity pump field in mode $a'$ is
\begin{equation}
\label{eq:extraPump}
V_p^{(+)}(t) = 
\begin{dcases}
\sqrt{\frac{\cP}{\hbar\omega_0}} e^{-\imag\omega_0 t}  & ; \quad \text{CW}\\
\sqrt{\frac{\cE}{2\pi \hbar\omega_0}} e^{-\imag\omega_0t} \int_{-\infty}^\infty d\Omega\, e^{-\imag\Omega t}A_p(\Omega) & ; \quad \text{pulsed}
\end{dcases},
\end{equation}
corresponding to the intracavity value of
\begin{equation}
\label{eq:intraPump}
F_p^{(+)}(z,t) = \begin{dcases}
\sqrt{\frac{4\tgamma_a v_g \cP}{\tgamma^2\hbar\omega_0 L}} e^{\imag(k_0z-\omega_0 t)}  & ; \quad \text{CW}\\
\sqrt{\frac{\tgamma_a v_g \cE}{2\pi \hbar\omega_0 L}} e^{\imag(k_0z-\omega_0t)} \int_{-\infty}^\infty \frac{d\Omega\,}{\frac{\tgamma}{2}-\imag\Omega} e^{-\imag\Omega t}A_p(\Omega)& ; \quad \text{pulsed}
\end{dcases}.
\end{equation}
Under these definitions $V_p^{(-)}(t)V_p^{(+)}(t) = \frac{\cP}{\hbar\omega_0}$ for the CW case (photons per second) and $\int dt\,V_p^{(-)}(t)V_p^{(+)}(t) = \frac{\cE}{\hbar\omega_0}$ for the pulsed case (photons per pulse).

The interaction Hamiltonian $\cH_I(t)$ for a single scalar field in a $\chi^{(3)}$ material is~\cite{Volkov2004}
\begin{equation}
\label{eq:Hamiltonian}
\cH_I(t) = -\frac{\chi^{(3)}}{4\epsilon_0^3n_0^8} \iiint_\mathcal{V} d^3 \mathbf{r} D(\br,t)D(\br,t)D(\br,t)D(\br,t),
\end{equation}
where in our case the displacement $D(\br,t)$ can be written as
\begin{equation}
\label{eq:Dfield}
D(\br,t) = \epsilon_0 n_0^2 \sqrt{\frac{\hbar\omega_0}{2\epsilon_0cn_0 S}} \left[\hF_s^{(+)}(z,t) + \hF_i^{(+)}(z,t) + F_p^{(+)}(z,t) + \text{h.c.}\right]
\end{equation}
with $\text{h.c.}$ denoting the Hermitian conjugate.
Plugging \cref{eq:Dfield} into \cref{eq:Hamiltonian}, keeping only the terms associated with SFWM, and integrating over the effective waveguide area $\iint dx dy = S$ returns
\begin{equation}
\label{eq:Hamiltonian2}
\cH_I(t) = \frac{n_2\hbar^2\omega_0^2}{cS} \int_0^L dz\, \left[F_p^{(+)}(z,t)\right]^2 \hF_s^{(-)}(z,t)\hF_i^{(-)}(z,t) + \text{h.c.},
\end{equation}
where we have neglected unimportant unimodular scale factors and made use of the standard nonlinear parameter definition $n_2=\frac{3\chi^{(3)}}{4\epsilon_0 c n_0^2}$~\cite{Boyd2020}.

Focusing on the first-order term in the power series, the quantum state in the interaction picture $\ket{\Psi} = \exp\left\{\frac{1}{\imag\hbar}\int_{-\infty}^t dt'\,\cH_I(t')\right\}\ket{\text{vac}}$ can be reduced to
\begin{equation}
\label{eq:Psi}
\ket{\Psi} = \frac{n_2\hbar\omega_0^2}{cS} \int_{-\infty}^\infty dt\int_0^L dz\, \left[F_p^{(+)}(z,t)\right]^2 \hF_s^{(-)}(z,t)\hF_i^{(-)}(z,t)\ket{\text{vac}},
\end{equation}
where $\ket{\text{vac}}$ denotes the vacuum state in signal and idler modes, unimodular scale factors have again been neglected, and the standard limit $t\rightarrow\infty$ has been applied; i.e., the interaction time exceeds all other characteristic timescales.

Our goal is to derive expressions for the extracted photons, namely the single-photon rates 
\begin{equation}
\label{eq:singles}
\begin{split}
R_s(t_s) & = \braket{\Psi|\hV_s^{(-)}(t_s)\hV_s^{(+)}(t_s)|\Psi} \\
R_i(t_s) & = \braket{\Psi|\hV_i^{(-)}(t_i)\hV_i^{(+)}(t_i)|\Psi} 
\end{split}
\end{equation}
and the second-order correlation function
\begin{equation}
\label{eq:G2}
G^{(2)}(t_s,t_i)  = \braket{\Psi|\hV_i^{(-)}(t_i)\hV_s^{(-)}(t_s)\hV_s^{(+)}(t_s)\hV_i^{(+)}(t_i)|\Psi},
\end{equation}
which for a pure two-photon state can be expressed in terms of a biphoton wavepacket~\cite{Shih2003}
\begin{equation}
\label{eq:wavepacket}
\psi(t_s,t_i)  = \braket{\text{vac}|\hV_s^{(+)}(t_s)\hV_i^{(+)}(t_i)|\Psi},
\end{equation}
as $G^{(2)}(t_s,t_i) = |\psi(t_s,t_i)|^2$.

\Crefrange{eq:intraFlux}{eq:wavepacket} outline a self-contained procedure for calculating the output photon rates from a single microring resonator under the assumption of energy-matched resonances (i.e., negligible dispersion) for arbitrary coupling conditions on the pump and biphoton, of which the three configurations in \cref{fig:setup} are special cases.
Our formalism is therefore significantly more general than previous interaction-picture microring models that focused on CW pumping and an all-pass geometry~\cite{Chen2011}. Moreover, as shown below, meticulous tracking of proportionality constants enables the estimation of absolute generation rates---a capability rare in microring SFWM models and typically associated with more tedious Heisenberg picture formulations~\cite{Chembo2016, Liu2024}. Of course, the perturbative underpinnings of \cref{eq:Psi} constrain our model's applicability to the low-flux (single-pair) regime. Yet we believe that doing so strikes a balance between simplicity and expressibility, encompassing the main design questions of photon-pair generation without additional complications. In the following two sections, we derive explicit formulas for each geometry and pumping configuration, obtain the coupling conditions which optimize single- and two-photon rates, and plot the behavior for some experimentally realistic examples.

\section{CW pumping}
\label{sec:CW}
\subsection{Theory}
Plugging \cref{eq:intraFlux} and the CW case of \cref{eq:intraPump} into \cref{eq:Psi} and simplifying yields
\begin{equation}
\label{eq:PsiCW}
\ket{\Psi} = \frac{4\tgamma_a}{\tgamma^2} \frac{n_2 v_g^2\omega_0\cP}{cSL} \sum_\nu\sum_\sigma \sqrt{\gamma_\nu\gamma_\sigma}\int_{-\infty}^\infty \frac{d\Omega_s }{\frac{\gamma^2}{4}+\Omega_s^2}\hnu_s^\dagger(\Omega_s)\hsigma_i^\dagger(-\Omega_s)\ket{\text{vac}},
\end{equation}
made possible by the energy-conservation and phase-matching conditions: $\omega_s+\omega_i = 2\omega_0$  and $k_s+k_i = 2k_0$. Conceptually, this expression highlights the fact that each photon from a generated pair can exit the cavity via any escape pathway---bus $a$, drop $b$, or loss $c$---independently and with probability proportional to the respective coupling rate $\gamma_a$, $\gamma_b$, or $\gamma_c$.

Straightforward calculations based on \cref{eq:extraFlux,eq:singles,eq:wavepacket} output the one-photon rates
\begin{equation}
\label{eq:singlesMu}
R_s = R_i = 32\frac{\tgamma_a^2\gamma_\mu}{\tgamma^4\gamma^2} \left(\frac{n_2v_g^2\omega_0\cP}{cSL}\right)^2 \quad
\end{equation}
and wavepacket
\begin{equation}
\label{eq:wavepacketMu}
\psi(\tau) = 4 \frac{\tgamma_a\gamma_\mu}{\tgamma^2\gamma} \frac{n_2v_g^2\omega_0\cP}{cSL} e^{-\gamma|\tau|/2},
\end{equation}
where $\tau=t_s-t_i$ and we continue to neglect unimodular factors. As expected for CW pumping, $R_s$ and $R_i$ are constant in time and $\psi$ depends only on the difference. The total pair extraction rate $R_{si}$ can be computed by integrating over the full wavepacket:
\begin{equation}
\label{eq:coincMu}
R_{si} =\int_{-\infty}^\infty d\tau\,G^{(2)}(\tau) = \int_{-\infty}^\infty d\tau\,|\psi(\tau)|^2 = 32 \frac{\tgamma_a^2\gamma_\mu^2}{\tgamma^4\gamma^3} \left(\frac{n_2v_g^2\omega_0\cP}{cSL}\right)^2.
\end{equation}

Interestingly, $R_{si}=\gamma_\mu R_s/\gamma=\gamma_\mu R_i/\gamma$, which possesses the intuitively satisfying interpretation that the total rate of output photon pairs is simply the single-photon rate multiplied by the probability the entangled partner escapes into the same output channel. This behavior grounds the tradeoff between total output rate and heralding efficiency (e.g., $R_{si}/R_i$) discussed previously~\cite{Vernon2016} and is reflected in the correlated term of the normalized $g^{(2)}$ function recently obtained for the all-pass, identical-coupling case~\cite{Liu2024}. To our knowledge, however, \crefrange{eq:singlesMu}{eq:coincMu} in their full generality 
have not been reported elsewhere in the literature.

\subsection{Identical pump-biphoton coupling}
\label{sec:CWhom}
Unless specific design measures are taken, such as interferometric couplers as depicted in \cref{fig:setup}(c), the pump, signal, and idler frequency bins typically experience the same coupling rates into and out of the microring. In this case, the distinction between pump and biphoton coupling can be removed by taking $\tgamma_\nu=\gamma_\nu\;\forall\nu\in\{a,b,c\}$ in \cref{eq:singlesMu,eq:coincMu}:
\begin{equation}
\label{eq:ratesHom}
\begin{split}
R_s = R_i = 32 \frac{\gamma_a^2\gamma_\mu}{\gamma^6} \left(\frac{n_2v_g^2\omega_0\cP}{cSL}\right)^2 \\
R_{si} = 32 \frac{\gamma_a^2\gamma_\mu^2}{\gamma^7} \left(\frac{n_2v_g^2\omega_0\cP}{cSL}\right)^2.
\end{split}
\end{equation}
Significantly, the one-photon results agree exactly with the maximum rate $R_\text{max}$ in Eq.~(79) of \cite{Chembo2016} after adjusting for differences in notation. For the purposes of optimization, we assume that the cavity loss rate $\gamma_c$ is fixed by material properties and device fabrication; the goal is to find $\gamma_a$ and $\gamma_b$ to maximize either the single-photon rate $R_s=R_i$ or two-photon rate $R_{si}$.

For the all-pass configuration [\cref{fig:setup}(a)], $\mu=a$ and $\gamma_b=0$; hence $\gamma=\gamma_a+\gamma_c$. Solving $\frac{\partial R_s}{\partial\gamma_a}=\frac{\partial R_i}{\partial\gamma_a}=0$ returns the maximum
\begin{equation}
\label{eq:maxSinglesHomAllPass}
\quad R_s^\text{max} = R_i^\text{max} = \frac{1}{2\gamma_c^3} \left(\frac{n_2v_g^2\omega_0\cP}{cSL}\right)^2 \quad\text{at}\quad \gamma_a=\gamma_c,
\end{equation}
which is precisely the critical coupling condition. On the other hand, the pair rate $R_{si}$ is maximized at a slightly higher coupling rate: 
\begin{equation}
\label{eq:maxCoincHomAllPass}
R_{si}^\text{max} = \frac{221,184}{823,543\gamma_c^3} \left(\frac{n_2v_g^2\omega_0\cP}{cSL}\right)^2\approx  \frac{0.2686}{\gamma_c^3} \left(\frac{n_2v_g^2\omega_0\cP}{cSL}\right)^2\quad\text{at}\quad \gamma_a=\frac{4\gamma_c}3.
\end{equation}
In other words, slight overcoupling $\gamma_a>\gamma_c$ maximizes the output pairs, which makes sense given the need to extract two photons rather than just one. 

In the add-drop configuration [\cref{fig:setup}(b)], $\mu=b$ and both $\gamma_a$ and $\gamma_b$ are nonzero. Optimizing both gives
\begin{equation}
\label{eq:maxHomAddDrop}
\begin{split}
R_s^\text{max} = R_i^\text{max} = \frac{2}{27\gamma_c^3} \left(\frac{n_2v_g^2\omega_0\cP}{cSL}\right)^2 \quad  & \text{at}\quad (\gamma_a,\gamma_b)=\left(\frac{2\gamma_c}{3},\frac{\gamma_c}{3}\right)\\
R_{si}^\text{max} = \frac{13,824}{823,543\gamma_c^3} \left(\frac{n_2v_g^2\omega_0\cP}{cSL}\right)^2\approx  \frac{0.0168}{\gamma_c^3} \left(\frac{n_2v_g^2\omega_0\cP}{cSL}\right)^2\quad &\text{at}\quad (\gamma_a,\gamma_b)=\left(\frac{2\gamma_c}{3},\frac{2\gamma_c}{3}\right).
\end{split}
\end{equation}
Interestingly, the maximum two-photon rate is significantly lower than for the all-pass geometry, with $R_{si}^\text{max}$ approximately $16\times$ smaller than than the value in \cref{eq:maxCoincHomAllPass}. From a design perspective, this finding points to the tradeoffs in selecting an add-drop configuration: while noise from the pump is reduced at the drop port (an advantage not captured in our idealized model), the presence of an additional coupling pathway $\gamma_a$ reduces the maximum possible pair extraction rate by more than an order of magnitude.

\subsection{Distinct pump-biphoton coupling}
\label{sec:CWhet}
As particularly pronounced in the add-drop findings above, the default situation of identical coupling rates for the pump and signal-idler constrains the design procedure by forcing a balance between the efficiency of trapping pump photons inside the microring and extracting signal and idler photons. In recent years, use of distinct coupling to lift this constraint in microring-based SFWM has grown significantly~\cite{Vernon2017, Tison2017, Liu2020, Burridge2023,  Borghi2024, Myilswamy2025b}. Typically leveraging asymmetric Mach--Zehnder interferometers, these designs allow for separately engineered coupling rates for the pump and signal-idler, which under pulsed pumping facilitate biphotons with Schmidt number~\cite{Law2000, Law2004, Bogdanov2006} below the fundamental limit $K=1.091$ set by identical coupling and a Gaussian pump spectrum~\cite{Vernon2017}.

Yet even in the CW case, distinct coupling offers advantages. Consider the setup in \cref{fig:setup}(c) tuned such that only the pump couples to the input waveguide ($\tgamma_a>0,\gamma_a=0$) and only the signal-idler couple to the drop waveguide ($\tgamma_b = 0,\gamma_b>0$). Assuming identical intracavity losses for the pump and biphoton ($\tgamma_c=\gamma_c$), \cref{eq:singlesMu,eq:coincMu} become
\begin{equation}
\label{eq:ratesHet}
\begin{split}
R_s = R_i = 32\frac{\tgamma_a^2\gamma_b}{(\tgamma_a+\gamma_c)^4(\gamma_b+\gamma_c)^2} \left(\frac{n_2v_g^2\omega_0\cP}{cSL}\right)^2\\
R_{si} = 32 \frac{\tgamma_a^2\gamma_b^2}{(\tgamma_a+\gamma_c) ^4(\gamma_b+\gamma_c)^3} \left(\frac{n_2v_g^2\omega_0\cP}{cSL}\right)^2,
\end{split}
\end{equation}
which are maximized by
\begin{equation}
\label{eq:maxHet}
\begin{split}
\quad R_s^\text{max} = R_i^\text{max} = \frac{1}{2\gamma_c^3} \left(\frac{n_2v_g^2\omega_0\cP}{cSL}\right)^2 \quad & \text{at}\quad (\tgamma_a,\gamma_b)=(\gamma_c,\gamma_c) \\
R_{si}^\text{max} = \frac{8}{27\gamma_c^3} \left(\frac{n_2v_g^2\omega_0\cP}{cSL}\right)^2\approx  \frac{0.2963}{\gamma_c^3} \left(\frac{n_2v_g^2\omega_0\cP}{cSL}\right)^2\quad & \text{at}\quad (\tgamma_a,\gamma_b)=(\gamma_c,2\gamma_c).
\end{split}
\end{equation}
The maximum one-photon rate obtains at critical coupling for both pump and biphoton simultaneously, equaling that of the all-pass, identical-coupling geometry [\cref{eq:maxSinglesHomAllPass}]. In other words, because critical coupling is best for both pump and generated photons, distinct coupling offers no advantage in terms of absolute \emph{single-photon} rates. On the other hand, the rate of extracted photon \emph{pairs} does benefit from distinct coupling: $R_{si}^\text{max}$ in  \cref{eq:maxHet} exceeds \cref{eq:maxCoincHomAllPass} by $\sim$1.1$\times$ and \cref{eq:maxHomAddDrop} by $\sim$18$\times$. In this way, distinct coupling obtains the best of both worlds---critical coupling for the pump ($\tgamma_a=\gamma_c$) and overcoupling for the biphoton ($\gamma_b=2\gamma_c$).

Naively, with the pump already maximally captured via critical coupling $\tgamma_a=\gamma_c$, one might expect the extreme overcoupled limit $\gamma_b\rightarrow\infty$ to optimize $R_{si}$, as this would ensure that every generated photon leaves in the drop port. Although optimum for heralding efficiency $R_{si}/R_i$, 
heavy overcoupling does not optimize the coincidence rate because of the importance of cavity enhancement on the vacuum modes. For even though intracavity biphotons do not explicitly stimulate photons in the perturbative single-pair regime, the vacuum modes can be viewed as doing so.

\subsection{Summary}
\label{sec:CWsummary}
\Cref{tab:CWsummary} summarizes our findings on optimizing coupling for both one-photon and two-photon rates in microring-based SFWM under CW pumping. The rate $R_0$ defined as
\begin{equation}
\label{eq:K}
R_0=\frac{1}{\gamma_c^3} \left(\frac{n_2v_g^2\omega_0\cP}{cSL}\right)^2
\end{equation}
encompasses all parameters independent of the coupling rates, which are assumed constant for comparison purposes. Looking across all geometries collectively, we note that:
\begin{enumerate}
\item The absolute one-photon rate is optimized by critical coupling of both the pump and biphoton (all-pass, identical; add-drop, distinct). 
\item The two-photon rate is optimized by critical coupling of the pump and $2\times$ overcoupling of the biphoton (add-drop, distinct). 
\item In any cases where pump and biphoton coupling cannot be controlled independently, the optimal condition is a tradeoff between these conflicting targets. 
\item The identical-coupling, add-drop case gives the lowest rates for both one- and two-photon outputs, due to additional effective loss channels---i.e., unwanted biphoton coupling to the input waveguide and unwanted pump coupling to the output waveguide.
\end{enumerate}

\begin{table}[tb!]
    \centering
    \footnotesize
    \begin{tabular}{|l|c|c|c|}
    \hline
  \textbf{CW} & All-pass, identical & Add-drop, identical & Add-drop, distinct \\\hline
   One-photon & $\gamma_a=\gamma_c$ &  $(\gamma_a,\gamma_b)=\left(\frac{2\gamma_c}{3},\frac{\gamma_c}{3}\right)$ & $(\tgamma_a,\gamma_b)=(\gamma_c,\gamma_c)$ \\
              & $\frac{1}{2}R_0=0.5R_0$    &  $\frac{2}{27}R_0\approx 0.0741R_0$ &  $\frac{1}{2}R_0=0.5R_0$  \\ \hline 
  Two-photon  & $\gamma_a=\frac{4\gamma_c}{3}$ & $(\gamma_a,\gamma_b)=\left(\frac{2\gamma_c}{3},\frac{2\gamma_c}{3}\right)$ & $(\tgamma_a,\gamma_b)=(\gamma_c,2\gamma_c)$ \\
              & $\frac{221,184}{823,543}R_0 \approx 0.2686R_0$ &  $\frac{13,824}{823,543}R_0\approx 0.0168R_0$ & $\frac{8}{27}R_0\approx 0.2963R_0$ \\ \hline
    \end{tabular}
    \caption{Summary of optimal coupling conditions for CW pumping of the microrings in \cref{fig:setup}.     The rate $R_0=(n_2v_g^2\omega_0\cP/cSL\gamma_c^{3/2})^2$ is fixed for all cases. Identical coupling is characterized by the conditions  $\tgamma_a=\gamma_a$ and $\tgamma_b=\gamma_b$, and distinct coupling independent pump and biphoton control ($\tgamma_a\neq\gamma_a,\tgamma_b\neq\gamma_b$). In the all-pass geometry, $\tgamma_b=\gamma_b=0$, and in the distinct case $\gamma_a=\tgamma_b=0$ (no signal-idler coupling to the input waveguide $a$, no pump coupling to the output waveguide $b$).}
    \label{tab:CWsummary}
\end{table}

\subsection{Simulations}
\label{sec:CWsims}
\begin{figure}[tb!]
\includegraphics[width=\textwidth]{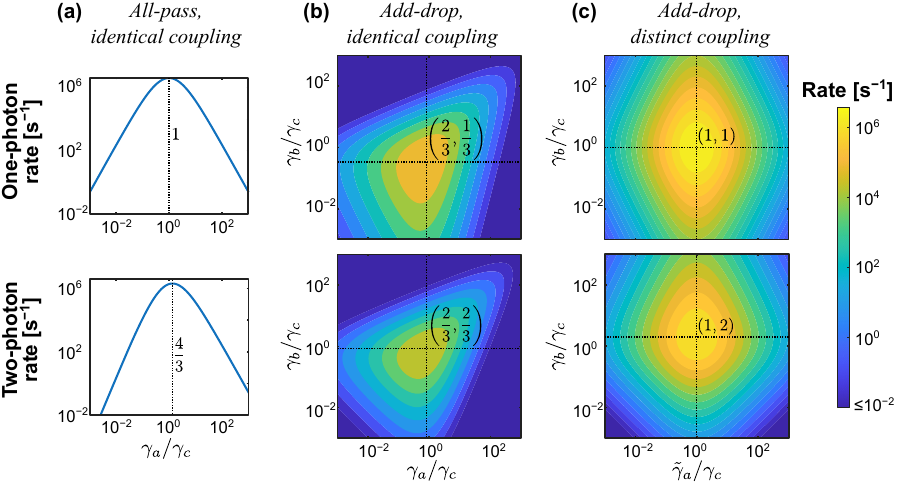}
\caption{Calculated one- and two-photon rates for CW pumping of an exemplar AlGaAs microring. (a) All-pass geometry with identical pump-biphoton coupling rates ($\tgamma_a=\gamma_a>0$, $\tgamma_b=\gamma_b=0$). (b) Add-drop geometry with identical coupling rates ($\tgamma_a=\gamma_a>0,\tgamma_b=\gamma_b>0$). (c) Add-drop geometry with distinct rates ($\tgamma_a>0,\gamma_b> 0,\gamma_a=\tgamma_b=0$). Dotted lines highlight the optimal coupling conditions (normalized to $\gamma_c$) for each case as summarized in \cref{tab:CWsummary}.}
\label{fig:CW}
\end{figure}

To explore the predictions of our model in a quantitative example, we consider an AlGaAs microring ($n_2=2.6\times10^{-17}$ m$^2$ W$^{-1}$~\cite{Chang2020}) pumped by $\cP=10$~$\upmu$W at $2\pi c/\omega_0 = 1550$~nm, with parameters inspired by \cite{Xie2020}: $\gamma_c/2\pi=71.1$~MHz, $v_g=8.57\times10^7$ m~s$^{-1}$, $S=0.330$~$\upmu$m$^2$, and $L/2\pi=143$~$\upmu$m---values that correspond to an intrinsic quality factor $Q_c = 2.72\times 10^6$ and free-spectral range of 95.4~GHz. \Cref{fig:CW} plots the rates for one- and two-photon outputs as functions of coupling: $\gamma_a$ only for the all-pass, identical-coupling case [\cref{fig:CW}(a)]; $\gamma_a$ and $\gamma_b$ for the add-drop, identical-coupling case [\cref{fig:CW}(b)]; and $\tgamma_a$ and $\gamma_b$ for the add-drop, distinct-coupling case [\cref{fig:CW}(c)]. The maximum used for the rate axis on all plots is $R_0/2 = 3.81\times10^6$~s$^{-1}$.

The appreciably lower rates for the add-drop geometry with identical coupling [\cref{fig:CW}(b)] matches what was found for the optimum specifically (\cref{tab:CWsummary}), and the interplay between pump and biphoton effects in this configuration creates noteworthy correlations between $\gamma_a$ and $\gamma_b$. Intuitively, if $\gamma_a$ is fixed at some value greater than the global optimum, $\gamma_b$ should be increased as well to ``pull'' the photons out of the drop port $b$ faster than they are extracted by the bus waveguide $a$.
In contrast, because of the factorization of pump and biphoton coupling effects in \cref{eq:ratesHet}, the add-drop geometry with distinct coupling [\cref{fig:CW}(c)] shows complete separability between the two rates; i.e., the optimal $\tgamma_b$ is completely independent of any fixed value of $\tgamma_a$, and vice versa. 
Perhaps one of the most interesting features is how forgiving the one- and two-photon rates are for nonoptimal coupling. For example, any $\gamma_a\in[0.30\gamma_c,3.4\gamma_c]$ keeps both the one-photon and two-photon rates to within 50\% of the peaks in \cref{fig:CW}(a). Overall, these results highlight not only how our model aids in the selection of nominal coupling values, but also how it can be used to predict the impact of fabrication tolerances on the absolute generation rates expected from a given device.

\section{Pulsed pumping}
\label{sec:pulsed}
\subsection{Theory}
For a pulsed pump, we take the intracavity pump field from the second case in \cref{eq:intraPump}, and a procedure similar to that preceding \cref{eq:PsiCW} yields
\begin{equation}
\label{eq:PsiPulsed}
\ket{\Psi} = \frac{\tgamma_a}{2\pi} \frac{n_2 v_g^2\omega_0\cE}{cSL} \sum_\nu\sum_\sigma \sqrt{\gamma_\nu\gamma_\sigma}\int_{-\infty}^\infty d\Omega_s \int_{-\infty}^\infty d\Omega_i  \frac{f_p(\Omega_s+\Omega_i)}{\left(\frac{\gamma}{2}-\imag\Omega_s\right)\left(\frac{\gamma}{2} -\imag\Omega_i \right)}\hnu_s^\dagger(\Omega_s)\hsigma_i^\dagger(\Omega_i)\ket{\text{vac}},
\end{equation}
where
\begin{equation}
\label{eq:fp}
f_p(\Omega_s+\Omega_i)= \int_{-\infty}^\infty d\Omega_p \frac{A_p(\Omega_p)A_p(\Omega_s+\Omega_i - \Omega_p)}{\left(\frac{\tgamma}{2}-\imag\Omega_p \right)\left[\frac{\tgamma}{2}-\imag(\Omega_s+\Omega_i-\Omega_p) \right]}
\end{equation}
denotes the standard effective pump lineshape function in SFWM~\cite{Vernon2017, Christensen2018, Myilswamy2023}. The probabilities per pulse of outputting a signal photon, $p_s = \int_{-\infty}^\infty dt\, \braket{\Psi|\hV_s^{(-)}(t)\hV_s^{(+)}(t)|\Psi}$, and idler photon, $p_i = \int_{-\infty}^\infty dt\, \braket{\Psi|\hV_i^{(-)}(t)\hV_i^{(+)}(t)|\Psi}$, are equal and given by
\begin{equation}
\label{eq:singlesPulsed}
p_s = p_i = \frac{\tgamma_a^2\gamma_\mu\gamma}{4\pi^2} \left(\frac{n_gv_g^2\omega_0\cE}{cSL}\right)^2 \int_{-\infty}^\infty d\Omega_s \int_{-\infty}^\infty d\Omega_i \frac{|f_p(\Omega_s+\Omega_i)|^2}{\left(\frac{\gamma^2}{4}+\Omega_s^2\right)\left(\frac{\gamma^2}{4}+\Omega_i^2 \right)},
\end{equation}
and the biphoton wavepacket possesses nontrivial dependencies on both time variables:
\begin{equation}
\label{eq:wavepacketPulsed}
\psi(t_s,t_i) =\frac{\tgamma_a\gamma_\mu}{4\pi^2} \frac{n_2v_g^2\omega_0\cE}{cSL} \int_{-\infty}^\infty d\Omega_s \int_{-\infty}^\infty d\Omega_i  \frac{f_p(\Omega_s+\Omega_i)}{\left(\frac{\gamma}{2}-\imag\Omega_s\right)\left(\frac{\gamma}{2} -\imag\Omega_i \right)}e^{-\imag(\Omega_s t_s + \Omega_i t_i)}.
\end{equation}

\Crefrange{eq:fp}{eq:wavepacketPulsed} apply to any pump spectrum and therefore can be exploited for numerical modeling of a variety of pump and coupling configurations. Yet, given the frequent use of pulsed pumps with spectra much broader than the pump resonance, for the remainder of the paper we examine the case of a flattop spectrum
\begin{equation}
\label{eq:flattop}
A_p(\Omega) = \begin{cases} \frac{1}{\sqrt{\Delta\Omega}} & ; |\Omega| < \frac{\Delta\Omega}{2} \\
0 & ; \text{otherwise}    
\end{cases},
\end{equation}
for which \cref{eq:fp} can be evaluated under the assumption $\Delta\Omega\gg\tgamma$ as:
\begin{equation}
\label{eq:fpBroadband}
f_p(\Omega_s+\Omega_i) = \frac{1}{\Delta\Omega} \int_{-\infty}^\infty \frac{d\Omega_p}{\left(\frac{\tgamma}{2}-\imag\Omega_p \right)\left[\frac{\tgamma}{2}-\imag(\Omega_s+\Omega_i-\Omega_p) \right]} 
= \frac{2\pi}{\Delta\Omega}\frac{1}{\tgamma-\imag(\Omega_s+\Omega_i)}.    
\end{equation}
Plugging this result into \cref{eq:wavepacketPulsed} and performing tedious but straightforward contour integration, we obtain the closed-form solution
\begin{equation}
\label{eq:wavepacketFinal}
\psi(t_s,t_i) =\frac{\tgamma_a\gamma_\mu}{\tgamma-\gamma} \frac{2\pi n_2v_g^2\omega_0\cE}{cSL\Delta\Omega} e^{-\gamma(t_s+t_i)/2} \left[1 - e^{-(\tgamma-\gamma)\min(t_s,t_i)} \right] u(t_s)u(t_i),
\end{equation}
where $u(t)$ denotes the Heaviside step function, the presence of which reflects causality: since the broadband pump is essentially a Dirac delta function $\delta(t)$ on the timescales of the microring, photons can only be produced for times $t>0$. Squaring and integrating over the full wavepacket returns the pair extraction probability per pulse:
\begin{equation}
\label{eq:probSIfinal}
p_{si} =\int_{-\infty}^\infty dt_s \int_{-\infty}^\infty dt_i |\psi(t_s,t_i)|^2 =  \frac{\tgamma_a^2\gamma_\mu^2}{\tgamma\gamma^2(\tgamma+\gamma)} \left(\frac{2\pi n_2v_g^2\omega_0\cE}{cSL\Delta\Omega}\right)^2,
\end{equation}
valid for pump bandwidth $\Delta\Omega\gg\tgamma$.

Similar integration  of \cref{eq:singlesPulsed} using $f_p(\cdot)$ as defined in \cref{eq:fpBroadband} yields
\begin{equation}
\label{eq:probSfinal}
p_s = p_i = \frac{\tgamma_a^2\gamma_\mu}{\tgamma\gamma(\tgamma+\gamma)} \left(\frac{2\pi n_2v_g^2\omega_0\cE}{cSL\Delta\Omega}\right)^2.
\end{equation}
As with the rates under CW pumping [\cref{eq:singlesMu,eq:coincMu}], the per-pulse probabilities confirm the intuitive expectation $p_{si} = \gamma_\mu p_s/\gamma = \gamma_\mu p_i/\gamma$; i.e., the probability of extracting a photon pair is simply the probability of extracting a single photon multiplied the probability $\gamma_\mu/\gamma$ its entangled partner also escapes into mode $\mu$.

\subsection{Optimal coupling conditions}
\label{sec:maxProb}
Because of the broadband pumping condition $\Delta\Omega\gg\tgamma$ underpinning \cref{eq:probSfinal,eq:probSIfinal}, selecting the coupling rates that maximize the extracted photon rates proves more nuanced than under CW pumping. Indeed, the optimal coupling rates for fixed spectral \emph{density} $\cE/\Delta\Omega$ are different than for a fixed pulse \emph{energy} $\cE$ in which $\Delta\Omega$ is scaled with $\tgamma$. And although either situation could reflect the constraints in a specific experimental context, for concreteness here we focus on fixed pulse energy as better aligned to the spirit of resource provision. For example, at a given repetition rate, carving a CW input into a desired pump pulse with bandwidth $\Delta\Omega$, followed by an erbium-doped fiber amplifier, holds $\cE$ fixed regardless of $\Delta\Omega$.

Accordingly, for the purposes of optimization, we take $\Delta\Omega$ proportional to the pump linewidth, i.e. $\Delta\Omega=B\tgamma$ for $B\gg1$. Under these conditions, the one- and two-photon extraction probabilities per pulse can be written in the compact forms
\begin{equation}
\begin{split}
\label{eq:probsCompact}
p_s = p_i =  \frac{\tgamma_a^2\gamma_\mu}{\tgamma^3\gamma(\tgamma+\gamma)}\gamma_c^2p_0\\
p_{si} = \frac{\tgamma_a^2\gamma_\mu^2}{\tgamma^3\gamma^2(\tgamma+\gamma)} \gamma_c^2p_0,
\end{split}
\end{equation}
where 
\begin{equation}
\label{eq:p0}
p_0 = \left(\frac{2\pi n_2 v_g^2\omega_0\cE}{cSLB\gamma_c} \right)^2
\end{equation}
is a dimensionless constant encompassing all terms independent of the coupling coefficients.

Following the same procedures as in \cref{sec:CWhom,sec:CWhet}, we obtain the optimal coupling conditions and corresponding generation probabilities summarized in \cref{tab:PulsedSummary}. 
While exact rational optima can be found in the two identical-coupling cases, the distinct-coupling formulas must be solved numerically.
Many trends prove quite similar to the CW cases in \cref{tab:CWsummary}. To within numerical precision, the add-drop geometry with distinct coupling reaches the same one-photon rate as the all-pass geometry, while slightly surpassing its two-photon maximum (by $\sim$1.14$\times$). Similarly, the add-drop geometry with identical coupling suffers the lowest rates of all three: $\sim$7$\times$ ($\sim$18$\times$) lower than the one-photon (two-photon) optimum in the other two cases. 

Interestingly, the presence of a continuum of pump frequencies---rather than a single peak at $\omega_0$---makes slight overcoupling for the pump optimal for maximizing the two-photon rate, even when it can be chosen completely independently of the biphoton coupling (add-drop, distinct). Heuristically, this can be understood through the intracavity flux  [\cref{eq:intraPump}], for which the energy at a given offset frequency $\Omega$ maximizes at $\tgamma_a=\sqrt{\gamma_c^2+4\Omega^2}$ (i.e., the critical coupling condition for frequency $\omega_0+\Omega$), suggesting an optimal value $\tgamma_a>\gamma_c$ when averaging over the full band. Finally, the condition $\gamma_a=2\gamma_c$ found for $p_{si}^\text{max}$ in the all-pass case with identical coupling matches that noted by \cite{Vernon2016} to optimize the extracted pair rate (called ``successful heralds'' there), thus offering another touchpoint of model validation~\cite{noteVernon2016}.

Continuing with the same AlGaAs microring and parameters considered in \cref{sec:CWsims}, but with pulse energy $\cE=1$~pJ and bandwidth factor $B=10$, we calculate the one- and two-photon probabilities in \cref{fig:pulsed} (top two rows). The shapes generally follow those of CW pumping (\cref{fig:CW}), although the distinct coupling case [\cref{fig:pulsed}(c)] now shows correlations between $\tgamma_a$ and $\gamma_c$, which follows mathematically from the $(\tgamma+\gamma)^{-1}$ factors in \cref{eq:probsCompact}. 

\subsection{Schmidt number}
\label{sec:schmidt}
So far, optimization of the broadband pulse cases has focused exclusively on generation probabilities. Yet historically, a primary motivation for broadband pumping in biphoton generation has been spectral purity or unentanglement~\cite{Grice2001,Mosley2008a}. 
For the pulsed wavepacket in \cref{eq:wavepacketFinal}, the degree of unentanglement can be quantified through the Schmidt decomposition~\cite{Law2000, Law2004, Bogdanov2006}
\begin{equation}
\label{eq:schmidtDecomp}
\psi(t_s,t_i) = \sum_{n=1}^\infty \sqrt{\lambda_n}\alpha_n(t_s)\beta_n(t_i),
\end{equation}
and the Schmidt number $K$ follows as $K = (\sum_n \lambda_n^2)/(\sum_n \lambda_n)$.
Perfect spectral purity obtains when $K=1$, meaning that heralding produces a pure single-photon state. 

\begin{table}[tb!]
    \footnotesize
    \centering
 \begin{tabular}{|l|c|c|c|}
    \hline
  \textbf{Broadband pulse} & All-pass, identical & Add-drop, identical & Add-drop, distinct \\\hline
   One-photon & $\gamma_a=\frac{3\gamma_c}{2}$  &  $(\gamma_a,\gamma_b)=\left(\gamma_c,\frac{\gamma_c}{2}\right)$ & $(\tgamma_a,\gamma_b)=(1.37\gamma_c,1.83\gamma_c)$ \\
              & $\frac{54}{3125}p_0\approx 0.0173p_0$    &  $\frac{8}{3125}p_0\approx 0.0026p_0$ &  $0.0173p_0$  \\ \hline 
  Two-photon  & $\gamma_a=2\gamma_c$ & $(\gamma_a,\gamma_b)=(\gamma_c,\gamma_c)$ & $(\tgamma_a,\gamma_b)=(1.46\gamma_c,3.17\gamma_c)$ \\
              &  $\frac{8}{729}p_0\approx 0.0110p_0$ &  $\frac{1}{1458}p_0\approx 0.0007p_0$ & $0.0125p_0$ \\ \hline
    \end{tabular}
    \caption{Summary of probability-maximizing coupling conditions for pumping by a broadband pulse of fixed energy. 
    The probability $p_0 = (2\pi n_2 v_g^2\omega_0\cE/cSLB\gamma_c)^2$ is held constant across  all cases. Identical coupling corresponds to equal pump and biphoton rates ($\tgamma_a=\gamma_a,\tgamma_b=\gamma_b$), and distinct to independent control ($\tgamma_a\neq\gamma_a,\tgamma_b\neq\gamma_b$). In the all-pass geometry, $\tgamma_b=\gamma_b=0$, and in the add-drop, distinct case, $\gamma_a=\tgamma_b=0$.}
    \label{tab:PulsedSummary}
\end{table}

However, $K=1$ can never be reached with simple broadband pumps in the identical coupling regime. Taking the limit $\tgamma\rightarrow\gamma$  in \cref{eq:wavepacketFinal} yields
\begin{equation}
\label{eq:wavepacketHomLim}
\lim_{\tgamma\rightarrow\gamma}\psi(t_s,t_i) = \tgamma_a\gamma_\mu \frac{2\pi n_2v_g^2\omega_0\cE}{cSL\Delta\Omega} e^{-\gamma(t_s+t_i)/2}\min(t_s,t_i) u(t_s)u(t_i),
\end{equation}
for which numerical calculations via the singular value decomposition~\cite{Bogdanov2006} return $K=1.091$.
Only by taking the limit $\tgamma\gg\gamma$ in \cref{eq:wavepacketFinal} is a fully separable biphoton obtained: 
\begin{equation}
\label{eq:wavepacketHetLim}
\lim_{\tgamma\gg\gamma}\psi(t_s,t_i) = \frac{\tgamma_a\gamma_\mu}{\tgamma} \frac{2\pi n_2v_g^2\omega_0\cE}{cSL\Delta\Omega} \underbrace{e^{-\gamma t_s/2} u(t_s)}_{\alpha_1(t_s)}  \underbrace{e^{-\gamma t_i/2}u(t_i)}_{\beta_1(t_i)},
\end{equation}
Because increasing $\tgamma$ without bound also suppresses generation efficiency, there exists an inherent tradeoff between temporal-spectral factorability and pair generation probability---at least for standard pulse shapes (no sharp spectral features on the order of the resonances themselves, like dual pulses~\cite{Christensen2018, Burridge2020}).

\begin{figure}[tb!]
\includegraphics[width=\textwidth]{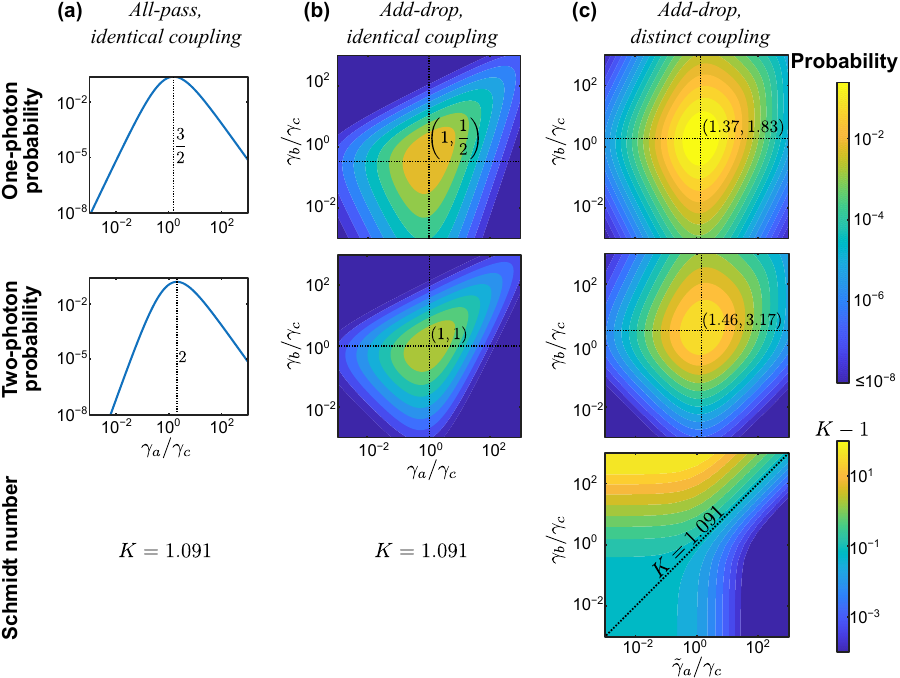}
\caption{Calculated rates and Schmidt numbers for broadband pumping of an exemplar AlGaAs microring. (a) All-pass geometry, identical pump-biphoton coupling ($\tgamma_a=\gamma_a>0$, $\tgamma_b=\gamma_b=0$). (b) Add-drop geometry, identical coupling rates ($\tgamma_a=\gamma_a>0,\tgamma_b=\gamma_b>0$). (c) Add-drop geometry, distinct rates ($\tgamma_a>0,\gamma_b> 0,\gamma_a=\tgamma_b=0$). Cases (a,b) correspond to the same Schmidt number $K=1.091$ for all coupling combinations. (See text for parameter values.) Dotted lines show the rate-maximizing coupling values (normalized to $\gamma_c$) for each plot.}
\label{fig:pulsed}
\end{figure}

To explore this tradeoff quantitatively, we include Schmidt number calculations for broadband-pumped biphoton states in the third row of \cref{fig:pulsed}. Because the pump is assumed much broader than the microring's linewidth---namely, $\Delta\Omega=10\tgamma$---the two identical-coupling cases have $K=1.091$ for all coupler settings. For the distinct-coupling configuration, we plot the calculated values as $K-1$ to highlight closeness to the ideal $K=1$ on a log-scale. Anything below the dashed line ($\gamma_b=\tgamma_a$) surpasses the spectral purity attainable with identical coupling, while the optimal one- and two-photon rates lie slightly above it; specifically, $K=1.119$ at $(\tgamma_a,\gamma_b)=(1.37\gamma_c,1.83\gamma_c)$ and $K=1.199$ at $(\tgamma_a,\gamma_b)=(1.46\gamma_c,3.17\gamma_c)$. Accordingly, whenever the desired Schmidt number is below these values, some reduction from the maximum rates must be tolerated in choosing the coupling, making the pairing $(\tgamma_a,\gamma_b)$ dependent on the ultimate use-case.

\section{Discussion}
Our model can be extended along a variety of directions. Multiple pairs of signal-idler frequency bins---i.e., a quantum frequency comb~\cite{Kues2019}---can be modeled via a sum over multiple $\omega_s$ and $\omega_i$ terms in \cref{eq:intraFlux,eq:extraFlux} then carried through the rest of the derivation. For a sufficiently broadband collection of bins such that GVD can no longer be neglected, the generation rates will reduce according to the frequency-dependent free-spectral range, typically expressed through a Taylor series of resonance frequencies $\omega_l = \omega_0 + \sum_{k\geq 1} \frac{\zeta_k}{k!} l^k$~\cite{Chembo2016, Kippenberg2018}. For example, under second-order dispersion ($\zeta_{k\geq 3} =0$), a slightly detuned CW pump at frequency $\omega_0+\Delta$, and signal (idler) photons with frequency $\omega_l+\Omega_s$ ($\omega_{-l}+\Omega_i$), energy conservation enforces $\Omega_s+\Omega_i = 2\Delta-\zeta_2 l^2$---a more complicated condition than $\Omega_s+\Omega_i=0$ that can nonetheless be handled analytically to model the suppression of pair generation for large $l$~\cite{Chembo2016}.
Similarly, the multi- and cascaded-micoring geometries growing in popularity for controllable frequency-bin state synthesis~\cite{Liscidini2019, Sabattoli2022, Clementi2023,Borghi2023,Alexander2025} present no major challenges to our method, for they again can be handled through the superposition of multiple intracavity and extracavity quantum fields.

Quantum states beyond two photons~\cite{Banic2024a, Banic2024b} prove more challenging, however, given our model's explicit restriction to the single-pair regime in \cref{eq:Psi}. Careful tracking of higher-order terms in the power series expansion of $\ket{\Psi} = \exp\left\{\frac{1}{\imag\hbar} \int_{-\infty}^t dt'\mathcal{H}_I(t') \right\}\ket{\text{vac}}$ could still be leveraged for quantitative insights into multiphoton effects, as shown in previous work on high-gain SPDC and SFWM~\cite{Quesada2014, Quesada2015, Quesada2022}. While definitely interesting to pursue, such complications begin to counteract the main motivation for the present work: deriving informative but simple results applicable to common microring pair-generation scenarios. 

Nevertheless, even with no additions to the current single-pair model, some multipair effects like accidental coincidences can be predicted. Using the standard and intuitive ``product-of-singles'' relationship for coincidences from uncorrelated phton paris~\cite{Eckart1938, Pearson2010, Truong2025}, the rate of accidentals for CW pumping follows as $R_\text{acc}=T_R R_s R_i$, with $T_R$ the coincidence window and $R_{s,i}$ defined in \cref{eq:singlesMu}; for pulsed pumping, the probability of an accidental coincidence per pulse can be approximated by$p_\text{acc} = p_s p_i$. Accordingly, our formalism can predict coincidence-to-accidental ratios (CARs) of microring sources.
Although this may prompt a desire to find the coupling rates that maximize CAR, it turns out 
that either ratio $R_{si}/R_\text{acc}$ or $p_{si}/p_\text{acc}$ approaches infinity in the limits of both extreme undercoupling ($\tilde{\gamma}_a\rightarrow 0$) and extreme overcoupling ($\tilde{\gamma}_a,\tilde{\gamma}_b\rightarrow\infty$)---two regimes corresponding to zero generation rates. Therefore unambiguous CAR optimization is not possible for finite rates: instead, just like nonresonant SFWM in bulk, CAR tuning is perhaps best achieved empirically by adjusting the pump power subject to some minimum tolerable detection rate in a given experiment.

\begin{backmatter}
\bmsection{Funding}
Oak Ridge National Laboratory, Laboratory Directed Research and Development (Heterogeneous Quantum Systems Initiative); Sandia National Laboratories, Laboratory Directed Research and Development (EPIQ, APT).

\bmsection{Acknowledgment}
This work was performed in part at Oak Ridge National Laboratory, operated by UT-Battelle for the U.S. Department of Energy under Contract No. DE-AC05-00OR22725. 

\bmsection{Disclosures}
The authors declare no conflicts of interest.

\bmsection{Data Availability Statement}
Data available upon reasonable request.

\end{backmatter}


\end{document}